\documentclass[a4paper,10pt,twoside]{cpc-hepnp}

\usepackage{multicol}
\usepackage{graphicx}
\usepackage{booktabs}
\usepackage{amssymb,bm,mathrsfs,bbm,amscd}
\usepackage[tbtags]{amsmath}
\usepackage{lastpage}
\usepackage{CJK}

\begin{document}
\begin{CJK*}{GBK}{song}

\fancyhead[c]{} \fancyfoot[C]{}

\footnotetext[0]{ }

\title{Melting temperature of heavy quarkonium with a
holographic potential up to sub-leading order}

\author{
\quad Zi-qiang Zhang$^{1,2}$\email{ziqiang85@139.com}%
\quad Yan Wu$^{1}$\email{yan.wu@cug.edu.cn}%
\quad De-fu Hou$^{2}$\email{hdf@iopp.ccnu.edu.cn}%
 }
 \maketitle

\address{%
$^1$ School of Mathematical and
Physics, China University of Geosciences(Wuhan), Wuhan 430074, China\\
$^2$ Key Laboratory of Quark and Lepton Physics (MOE),
Central China Normal University, Wuhan 430079, China\\
}

\begin{abstract}
A calculation of the melting temperatures of heavy quarkonium
states with the holographic potential was introduced in a previous
work. In this paper, we consider the holographic potential at
sub-leading order, which permits finite coupling corrections to be
taken into account. It is found that this correction lowers the
dissociation temperatures of heavy quarkonium.
\end{abstract}



\footnotetext[0]{\hspace*{-3mm}\raisebox{0.3ex}{}}%

\begin{multicols}{2}

\section{Introduction}
Heavy quarkonium dissociation is an important signal of the
formation of Quark-Gluon Plasma(QGP)in heavy ion collisions at
RHIC and LHC. However, much experiment data indicates that the QGP
is strongly coupled. Thus, the study of heavy quarkonium and its
dissociation requires non-peturbative techniques, such as Lattice
QCD and potential models\cite{AM}. The lattice simulation of the
quark-antiquark potential and the spectral density of hadronic
correlators yield a consistent picture of quarkonium dissociation
as well as the numerical value $T_d$. On the other hand, heavy
quarkonium dissociation can be studied within potential models,
for instance the energy levels and the dissociation temperature
can be carried out with the aid of a non-relativistic Schrodinger
equation with a temperature dependent effective potential when we
neglect the velocity($v\ll1$) of the constituent
quarks\cite{EV,WM,AM1}, and if we consider the relativistical
correction, the two-body Dirac equation can be employed\cite{WY}.
The holographic potential addressed in this paper is one example
of a potential model.

Holographic potential at finite temperature at strong coupling
relies on the AdS/CFT duality which can explore the strongly
coupled $\mathcal N=4$ supersymmetric Yang-Mills(SYM) plasma
through the correspondence between the type IIB superstring theory
formulated on AdS$_5\times S^5$ and $\mathcal N=4$ SYM in four
dimensions\cite{Maldacena:1997re,Gubser:1998bc,Witten:1998qj,MadalcenaReview,Maldacena:1998im}.

In a previous work \cite{DF}, the melting temperatures of heavy
quarkonium states were studied with the holographic potential. In
this paper we consider the holographic potential including its
sub-leading order and obtain the correction to the melting
temperatures.

The paper is organized as follows. In the next section, we will
present our strategy for computation. The sub-leading order of the
holographic potential and its corrections to the dissociation
temperatures will be discussed in section 3 and section 4. Section
5 concludes the paper.

\section{Setup}

Heavy quarkonium, $J/\psi$ or $\Upsilon$, can be modelled as a
non-relativistic bound state of a heavy quark and its
antiparticle, and the wave function of their relative motion
satisfies the Schrodinger equation
\begin{equation}
[-\frac{1}{2\mu}\nabla^2+U(r,T)]\psi=-E(T)\psi
\end{equation}
$E(T)$ is the binding energy of the bound state and $U(r,T)$ is
related to the free energy $F(r,T)$ by
\begin{equation}
U(r,T)=-T^2[\frac{\partial}{\partial T}(\frac{F(r,T)}{T})]_r
\end{equation}
$F(r,T)$ can be extracted from the Wilson loop operator between a
static pair of $q\bar{q}$
\begin{equation}
e^{-\frac{1}{T}F(r,T)}=\frac{tr<W^\dag(L_+)W(L_-)>}{tr<W^\dag(L_+)><W(L_-)>}
\end{equation}
where $L_\pm$ denotes the Wilson line running in the Euclidean
time direction at spatial coordinates $(0,0,\pm\frac{1}{2}r)$ and
is closed by the periodicity $\beta=\frac{1}{T}$ and
\begin{equation}
W(L_\pm)=Pe^{-i\oint_{L_{\pm}} dx^\mu A_\mu(x)}
\end{equation}
with $A_\mu$ the gauge potential and the symbol $P$ enforcing the
path ordering along the loop $C$. The thermal expectation value
$<W(C)>$ can be measured for QCD on a lattice and the heavy quark
potential is defined with F-ansatz or U-ansatz.

The holographic principle places $L_\pm$ on the
boundary$(z\rightarrow0)$of the Schwarzschild-AdS$_5\times S^5$,
whose metric can be written as
\begin{equation}
ds^2=\pi^2T^2z^2(fdt^2+d\vec{x}^2)+\frac{1}{\pi^2T^2z^2f}dz^2
\end{equation}
where $f=1-\frac{1}{z^4}$, $d\vec{x}^2=dx_1^2+dx_2^2+dx_3^2$ with
$x_1=x_2=0$ and $x_3$ a function of z.

In the case of $\mathcal N=4$ SYM, the AdS/CFT duality relates the
Wilson loop expectation value to the path integral of the
string-sigma action developed in \cite{Metsaev:1998it} of the
worldsheet in the AdS$_5\times S^5$ bulk.

To leading order of the strong coupling, the path integral is
given by its classical limit, which is the minimum area of the
world sheet
\begin{equation}
F(r,T)=-\frac{4\pi^2}{\Gamma^4\left(\frac{1}{4}\right)}\frac{\sqrt{\lambda}}{r}
{\rm min}[g_0(rT),0]\label{free},
\end{equation}
with
\begin{eqnarray}
-\frac{4\pi^2}{\Gamma^4\left(\frac{1}{4}\right)}\frac{\sqrt{\lambda}}{r}
g_0(rT)&=&\frac{1}{\pi\alpha^\prime}\Big[\int_0^{z_0}dz
\left(\frac{\sqrt{f}z_0^2}{z^2\sqrt{z_0^4-z^4}}
-\frac{1}{z^2}\right)\nonumber\\
&-&\int_{z_0}^{z_h}\frac{dz}{z^2}\Big],
\end{eqnarray}
where $g_0(rT)$ is a monotonically decreasing function with
$g_0(0)=1$, $g_0(r_0T)=0$ and $r_0$ is the screening length. If we
introduce a dimensionless radial coordinate $\rho=\pi T r$, we
have
\begin{equation}
g_0(\rho)=1-\frac{\rho}{\rho_0}
\end{equation}
with $\rho_0=0.7359$.

The melting temperatures of heavy quarkonium states with the
leading order potential related to (\ref{free}) were discussed
in\cite{DF}.

\section{The holographic potential model}
Now we add the sub-leading order term to the holographic potential
and explore its contribution.

As was shown in Ref\cite{ZQ}, the strong coupling expansion of
$F(r,T)$ at large $\lambda$ can be written as
\begin{equation}
F(r,T)=-\frac{4\pi^2}{\Gamma^4\left(\frac{1}{4}\right)}\frac{\sqrt{\lambda}}{r}
{\rm min}\Big[g_0(rT)-\frac{1.3346
g_1(rT)}{\sqrt{\lambda}}+O(\frac{1}{\lambda}),0\Big]
\label{expansion}
\end{equation}
where $g_1(rT)$ is a monotonically decreasing function, which
reaches 0.92 at $r_0$.

Likewise, we use the dimensionless radial coordinate $\rho=\pi T
r$, and we find
\begin{equation}
F(r,T)=-\frac{\alpha}{r}\phi(\rho)\theta(\rho_1-\rho),
\end{equation}
where
$\alpha=\frac{4\pi^2}{\Gamma^4\left(\frac{1}{4}\right)}\sqrt{\lambda}\simeq0.2285\sqrt{\lambda}$,
 $\rho_1$ is determined by $\phi(\rho_1)=0$.

The analytical small $\rho$ expansion and numerical results of
$\phi(\rho)$ both suggest
\begin{eqnarray}
\phi(\rho)&=&g_0(\rho)-\frac{1.3346 g_1(\rho)}{\sqrt{\lambda}}\nonumber\\
&\simeq&1-\frac{\rho}{\rho_0}-\frac{1.3346g_1(\rho)}{\sqrt{\lambda}}
.
\end{eqnarray}

As the temperature correction to the sub-leading term of the heavy
quark potential is small, or in other words $g_1(\rho)$ decreases
monotonically from 1 to 0.92 as $\rho\in(0,\rho_0)$, we can fit
$g_1(\rho)=1-0.11\rho$. This yields
\begin{equation}
\phi(\rho)
=1-\frac{\rho}{\rho_0}-\frac{1.3346(1-0.11\rho)}{\sqrt{\lambda}}
,\label{rou}
\end{equation}

To proceed, we define the dissociation temperature $T_d^\prime$ as
the temperature when the bound energy $E(T_d^\prime)$ becomes
zero, and the corresponding Schrodinger equation reduces to
\begin{equation}
\frac{d^2R}{d\rho^2}+\frac{2}{\rho}\frac{dR}{d\rho}-[\frac{l(l+1)}{\rho^2}+U]R=0\label{equation}
\end{equation}
with $U=\frac{mV_{eff}}{\pi^2T^2}$.

Actually, one should consider the exact holographic potential, but
it has been found in \cite{WY} that the comparison with the
dissociation temperature obtained from (\ref{rou}) is very close
to that from the exact holographic potential. So we stay with the
truncated Coulomb potential for the rest of the paper. Here we
consider the U-ansatz, where we have
\begin{eqnarray}
U=-\frac{\eta^2}{\rho_1\rho}[\phi(\rho)-\rho(\frac{d\phi}{d\rho})]\theta(\rho_1-\rho)\label{u}
\end{eqnarray}
with
\begin{equation}
\eta=\sqrt{\frac{\alpha\rho_1m}{\pi T}}\label{eta}.
\end{equation}
It follows from (\ref{rou}) and (\ref{u}) that
\begin{eqnarray}
U=-\frac{\eta^2}{\rho_1\rho}(1-\frac{1.3346}{\sqrt{\lambda}})\theta(\rho_1-\rho)
\label{u1}
\end{eqnarray}

On writing
\begin{equation}
\eta_x=\eta\sqrt{1-\frac{1.3346}{\sqrt{\lambda}}}\label{etax},
\end{equation}
we have
\begin{eqnarray}
U=-\frac{\eta_x^2}{\rho_1\rho}\theta(\rho_1-\rho) \label{u1}.
\end{eqnarray}

Substituting (\ref{u1}) into (\ref{equation}) one gets
\begin{eqnarray}
R(r)&=&\frac{1}{\sqrt{\rho}}J_{2l+1}(2\eta_x\sqrt{\frac{\rho}{\rho_1}}),\qquad
\rho\leq\rho_1,\nonumber\\
 R(r)&=&const.\rho^{-l-1},\qquad\rho>\rho_1\label{R}
\end{eqnarray}
with $J_n(x)$ the Bessel function. Then the threshold $\eta_x$ can
be related to the matching condition at $\rho=\rho_1$
\begin{equation}
\frac{d}{d\rho}(\rho^{l+1}R(r))|_{\rho=\rho_1^-}=0,
\end{equation}
this yields the correspond secular equation for $\eta_x$
\begin{equation}
2l+1-\eta_x\frac{J_{2l+2}(2\eta_x)}{J_{2l+1}(2\eta_x)}=0
\label{eta1}.
\end{equation}

Knowing the values of $l$, $\eta_x$ can be calculated from
(\ref{eta1}).

Finally, the melting temperature of heavy quarkonium with a
holographic potential up to sub-leading order can be obtained
according to:
\begin{equation}
T_d^\prime=\frac{\alpha\rho_1m}{\pi
\eta^2}=\frac{\alpha\rho_1m}{\pi
\eta_x^2}(1-\frac{1.3346}{\sqrt{\lambda}})\label{td}.
\end{equation}

\section{Results}

Now we discuss our results. At first, we take
$\lambda\rightarrow\infty$ in (\ref{rou}) and (\ref{td}), which
leads to the leading order case
\begin{equation}
T_d=\frac{\alpha\rho_0m}{\pi \eta^2},\label{td1}
\end{equation}
where $T_d$ is the melting temperature of heavy quarkonium states
with the leading order potential. The numerical results for $T_d$
of $J/\Psi$ and $\Upsilon$ are presented in Table 1, where we have
chosen $m=1.65GeV,4.85GeV$ for c and b quarks.
\begin{center}
\tabcaption{ \label{tab1}  $T_d$ in MeV for $J/\Psi$ and
$\Upsilon$ under the holographic potential.} \footnotesize
\begin{tabular*}{80mm}{c@{\extracolsep{\fill}}ccc}
\toprule  & $T_d(\lambda=5.5)$   & $T_d(\lambda=6\pi)$   \\
\hline
$J/\Psi$(1s) &143 &265 \\
$J/\Psi$(2s) &27 &50 \\
$J/\Psi$(1p) &31 &58 \\
$\Upsilon$(1s) &421 &780 \\
$\Upsilon$(2s) &80 &148 \\
$\Upsilon$(1p) &92 &171 \\
\bottomrule
\end{tabular*}
\end{center}

Then we consider sub-leading order correction. For comparison,
here we we show the curve about $T_d'/T_d$ vs $\lambda$ in Fig.1.
Note that owing to the $\lambda$ correction to the holographic
potential, $T_d'$ is smaller than $T_d$. With the typical interval
$5.5<\lambda<6\pi$\cite{GB}, we find this correction gives rise to
a 47\% reduction of $T_d$ when $\lambda=6\pi$, and it will
increase as $\lambda$ becomes smaller. One may doubt this result
because with small values of $\lambda$ the corrections term in
(\ref{expansion}) and (\ref{td}) will be very large so that these
corrections are meaningless. However, this consideration is
unnecessary since the strong coupling expansion of the potential
relies on the assumption that the $\lambda$ is large. Indeed, it
appears that this correction will vanish for
$\lambda\rightarrow\infty$ as it must, since the sub-leading order
to the potential vanishes in that limit.
\begin{center}
\includegraphics[width=9cm]{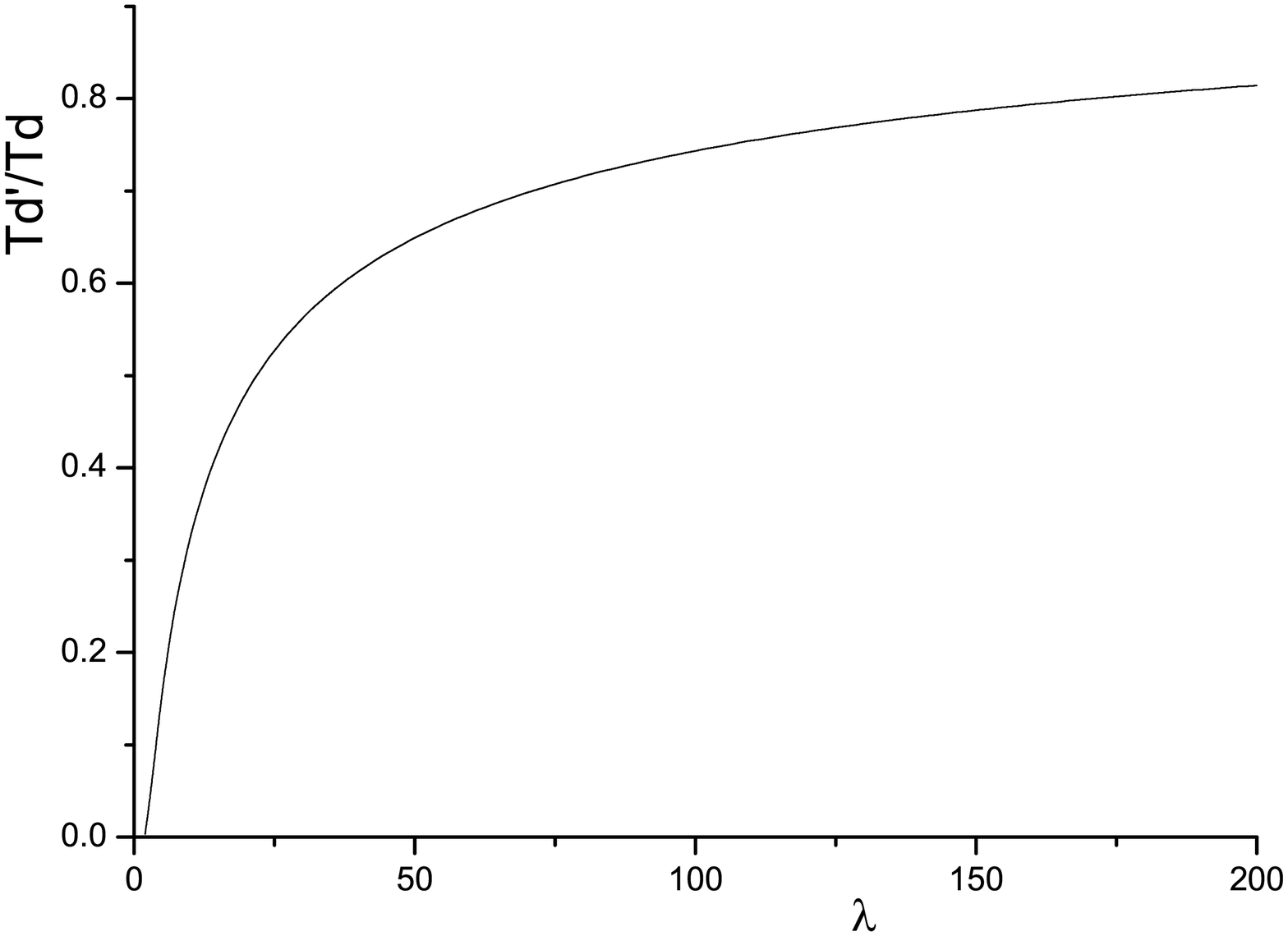}
\figcaption{\label{fig1} $T_d'/T_d$ vs $\lambda$. Melting
temperature with sub-leading order potential over its counterpart
with potential versus $\lambda$.}
\end{center}

\section{Conclusion}
In this paper, we have investigated the melting temperatures
studied with the truncated holographic potential and we consider
the holographic potential with its sub-leading order, which
permits to take into account finite coupling corrections. It is
found that this correction becomes smaller as $\lambda$ increases.
With this corrections, the dissociation temperatures of heavy
quarkonium are lowered, leaving the corrected values further below
the lattice result. This disagreement can be attributed to several
reasons. Firstly, the short screening length $r_0\simeq0.25fm$ at
$T=200MeV$ of the $AdS/CFT$ potential and sharp cutoff nature of
the screening. Secondly, one should take into account the
different number of degrees of freedom in $N_c=3$ SYM and 3 flavor
QCD; a similar problem has been explained in the calculation of
the jet quenching parameter beyond $AdS/CFT$ correspondence
in\cite{KM}, where they have matched the corresponding entropy
density to obtain $T^3\simeq3T^3_{SYM}$. Moreover, one should also
bear in mind that the particles of $\mathcal N=4$ SYM are quite
different to that of QCD. In particular it does not include
particles in the fundamental representation, but only in the
adjoint representation.

To summarize, the melting temperature studied in this work relies
on the holographic quark potential. However, in gauge-gravity
duality, heavy quark potential at finite temperature is usually
calculated with the pure AdS background, and the potential also
does not contain any confining term in the deconfined phase. This
has led some authors to consider the potential closer to QCD, for
instance heavy quark potential in strongly-coupled $\mathcal N=4$
SYM in a magnetic field\cite{RP} and with some deformed $AdS_5$
model\cite{HM}. Applying these rectified potentials, one can
obtain the melting temperature as well.

However, we should admit that there are some shortcomings in our
work. Firstly, our updated results do not seem to be close to
"reality"; this may depend on the model. Secondly, recently many
authors have suggested that the potential at non-zero temperature
is complex\cite{YB,MA}: the real part is neither the free energy
nor the internal energy, and the imaginary potential plays an
important role in setting the dissociation temperature. This
subject is very interesting, and we hope to do some future work in
this regard.

\section{Acknowledgments}
We would like to thank Prof. Hai-cang Ren for useful discussions.
The research of Zi-qiang Zhang is partly supported by the
Fundamental Research Funds for the Central Universities and the
QLPL under grant Nos. QLPL201408. The work of De-fu Hou is
supported in part by NSFC under Grant Nos. 11375070, 11135011 and
11221504.

\begin{subequations}
\renewcommand{\theequation}{A\arabic{equation}}

\end{subequations}

\end{multicols}

\vspace{-1mm}
\centerline{\rule{80mm}{0.1pt}}
\vspace{2mm}

\begin{multicols}{2}

\end{multicols}

\clearpage

\end{CJK*}

\begin{thebibliography}{90}
\bibitem{AM}
Agnes Mocsy, [hep-ph/0811.0337v1].

\bibitem{EV}
E.V.Shuryak and I.Zahed,Phys.Rev.{\bf D70}, (2004) 054507.

\bibitem{WM}
W.M.Alberico.et al, Phys.Rev.{\bf D72}, (2005) 114011.

\bibitem{AM1}
Agnes Mocsy,.et al, Phys.Rev.{\bf D77}, (2008) 014501.

\bibitem{WY}
Yan Wu, Defu Hou and Hai-cang Ren,Phys.Rev. {\bf C87}, (2013)
025023.

\bibitem{Maldacena:1997re}
J.~M.~Maldacena,, Adv.\ Theor.\ Math.\ Phys.\  {\bf 2}, 231 (1998)
[Int.\ J.\ Theor.\ Phys.\  {\bf 38}, 1113 (1999)].

\bibitem{Gubser:1998bc}
S.~S.~Gubser, I.~R.~Klebanov and A.~M.~Polyakov, Phys.\ Lett.\  B
{\bf 428}, 105 (1998) [hep-th/9802109].

\bibitem{Witten:1998qj}
E.~Witten,  {\bf 2}, 253 (1998) [hep-th/9802150].

\bibitem{MadalcenaReview}
O. Aharony, et al, Phys. Rept. {\bf 323}, 183 (2000).

\bibitem{Maldacena:1998im}
J.~M.~Maldacena, Phys.\ Rev.\ Lett.\  {\bf 80}, 4859 (1998).

\bibitem{DF}
De-fu Hou, Hai-cang Ren, JHEP {\bf 01}, (2008) 029.

\bibitem{Metsaev:1998it}
R.~R.~Metsaev and A.~A.~Tseytlin, Nucl.\ Phys.\  B {\bf 533}, 109
 (1998) .

\bibitem{ZQ}
Zi-qiang Zhang, De-fu Hou, Hai-cang Ren and Lei Yin JHEP {\bf 07},
 (2011) 035.

\bibitem{GB}
S.Gubser, Phys.Rev. {\bf D76}, (2007) 126003.

\bibitem{KM}
Karen M. Burke, et al, Phys. Rev. {\bf C90}, (2014) 014909.

\bibitem{RP}
R. Rougemont, et al, [hep-th/1409.0556v1].

\bibitem{HM}
Song He, Mei Huang and Qi-shu Yan, Prog.Theor.Phys.Suppl. {\bf
186}, (2010) 504-509.

\bibitem{YB}
Y.Burnier, A.Rothkopf, [hep-ph/1506.08684].

\bibitem{MA}
M. A. Escobedo, [hep-ph/1401.4892].

\end{thebibliography}
\end{document}